\documentclass[twocolumn]{aastex63}

\usepackage{amsmath}
\usepackage{graphicx}
\usepackage{bm}
\usepackage{booktabs}
\usepackage{apjfonts}
\usepackage{xcolor}
\usepackage{appendix}
\usepackage{url}

\newcommand{\mr}[1]{\mathrm{#1}}

\newcommand{\lum}{erg\,s\ensuremath{^{-1}}}
\newcommand{\msun}{\ensuremath{M_{\odot}}}

\shorttitle{EP240222a Five-Stage Model}
\shortauthors{Li et al.}

\begin{document}

\title{\Large Inefficient Circularization, Delayed Stream-Disk Interaction and Reprocessing: A Five-Stage Model for the Intermediate-Mass Black Hole Tidal Disruption Event EP240222a}

\author[0009-0007-3464-417X]{Wenkai Li}
\affiliation{Department of Astronomy, University of Science and Technology of China, Hefei 230026, China; liwenkai@mail.ustc.edu.cn, jnac@ustc.edu.cn}
\affiliation{School of Astronomy and Space Sciences,
University of Science and Technology of China, Hefei 230026, China}
\affiliation{Steward Observatory, University of Arizona, 933 North Cherry Avenue, Tucson, AZ 85721, USA}

\author[0000-0002-7152-3621]{Ning Jiang}
\affiliation{Department of Astronomy, University of Science and Technology of China, Hefei 230026, China; liwenkai@mail.ustc.edu.cn, jnac@ustc.edu.cn}
\affiliation{School of Astronomy and Space Sciences,
University of Science and Technology of China, Hefei 230026, China}

\author[0000-0002-1517-6792]{Tinggui Wang}
\affiliation{Department of Astronomy, University of Science and Technology of China, Hefei 230026, China; liwenkai@mail.ustc.edu.cn, jnac@ustc.edu.cn}
\affiliation{School of Astronomy and Space Sciences,
University of Science and Technology of China, Hefei 230026, China}


\author[0000-0001-5012-2362]{Rongfeng Shen}
\affiliation{School of Physics and Astronomy, Sun Yat-sen University, Zhuhai 519082, China}
\affiliation{CSST Science Center for the Guangdong-Hong Kong-Macau Greater Bay Area, Sun Yat-sen University, Zhuhai 519082, China}

\author[0000-0001-8319-6034]{Erlin Qiao}
\affiliation{National Astronomical Observatories, Chinese Academy of Sciences, Beijing 100101, China}
\affiliation{School of Astronomy and Space Science, University of Chinese Academy of Sciences, Beijing 100049, China}

\author[0000-0002-9589-5235]{Lixin Dai}
\affiliation{Department of Physics, The University of Hong Kong, Pokfulam Road, Hong Kong, China}

\author[0009-0007-1153-8112]{Di Luo}
\affiliation{Department of Astronomy, University of Illinois at Urbana-Champaign, Urbana, IL 61801, USA}

\author[0000-0002-4562-7179]{Dongyue Li}
\affiliation{National Astronomical Observatories, Chinese Academy of Sciences, Beijing 100101, China}

\author[0000-0002-2006-1615]{Chichuan Jin}
\affiliation{National Astronomical Observatories, Chinese Academy of Sciences, Beijing 100101, China}
\affiliation{School of Astronomy and Space Science, University of Chinese Academy of Sciences, Beijing 100049, China}
\affiliation{Institute for Frontier in Astronomy and Astrophysics, Beijing Normal University, Beijing 102206, China}

\author[0000-0003-3824-9496]{Jiazheng Zhu}
\affiliation{Department of Astronomy, University of Science and Technology of China, Hefei 230026, China; liwenkai@mail.ustc.edu.cn, jnac@ustc.edu.cn}
\affiliation{School of Astronomy and Space Sciences,
University of Science and Technology of China, Hefei 230026, China}


\begin{abstract}
EP240222a is the first intermediate-mass black hole (IMBH) tidal disruption event (TDE) captured in real-time with multi-wavelength observations and spectroscopic confirmation. However, its light curves deviate substantially from previous theoretical expectations. Motivated by these unique features, we have developed a novel model that successfully reproduces its peculiar evolution. Our model delineates five stages: (1) Initial Stage of inefficient circularization; (2) Slow-Rising Stage with a faint X-ray precursor disk fed by successive self-crossings; (3) Fast-Rising Stage, where delayed stream-disk interaction at momentum flux matching drives a sharp luminosity rise; (4) Plateau Stage with super-Eddington accretion, outflow, reprocessing, and a clear polar line-of-sight; and (5) Decline Stage of sub-Eddington accretion and ongoing reprocessing. Our fit indicates the disruption of a $M_* \approx 0.4~M_\odot$ main-sequence (MS) star with a penetration factor $\beta \approx 1.0$. Our model, which incorporates key TDE processes, establishes EP240222a-like light curves as typical IMBH-TDE signatures. The distinctive identifier is a slow rise in X-rays and a corresponding slow rise/quasi-plateau in the UV/optical, followed by a brighter, super-Eddington plateau in both bands, though other forms exist, such as the rapid rise from white dwarf (WD) disruptions over minutes to days.
\end{abstract}


\section{Introduction}

When a star ventures too close to a supermassive black hole (SMBH) at the center of a galaxy, it can be torn apart if the pericenter of its orbit lies within the tidal disruption radius. Such an occurrence is known as a tidal disruption event (TDE). These events are followed by multi-wavelength electromagnetic emission, offering a powerful means for detecting otherwise elusive dormant SMBHs~\citep{Rees1988}. Over the past decade, the advance of time-domain surveys has led to a growing number of observed TDEs~\citep{Gezari2021}, gradually enabling demographic studies of local SMBHs~\citep{vV2018,Yao2023}. 

Moreover, TDEs are a valuable tool to detect intermediate-mass black holes (IMBHs). With masses of $10^2$–$10^5~M_\odot$, IMBHs bridge the gap between SMBHs in galactic nuclei and stellar-mass black holes in binary systems and provide critical insights into the mass and abundance of SMBH seeds in the early universe (see \citealt{Greene2020} as a review). Currently, most IMBH candidates are identified via active galactic nucleus (AGN) features, with masses being derived either directly from AGN emission (e.g., \citealt{Greene2004,Greene2007,Dong2012,Reines2013,Liu2018,Liu2025}) or via a scaling relation based on their dwarf host galaxies (e.g., \citealt{Reines2015,Mezcua2018,Ward2022}). TDEs by IMBHs thus offer a unique probe of their inactive counterparts. This is compelling because theoretical TDE rates are higher at the low-mass end, i.e. dwarf galaxies~\citep{Wang2004,Stone2016}, and a substantial population of IMBHs may reside in off-nuclear regions surrounded by scarce accretable gas, unlike those found in galactic nuclei.

The unique scientific value of IMBH-TDEs has prompted an increasing number of searches for IMBHs in the vast datasets from time-domain surveys. Several unusual phenomena have been attributed to candidate IMBH-TDEs, such as the off-center soft X-ray outburst 3XMM J215022.4–055108 in a massive galaxy~\citep{Lin2018, Zhang2025} and the fast-rising nuclear transient AT2020neh in a dwarf galaxy~\citep{Angus2022}. However, these interpretations remained inconclusive due to a lack of key TDE signatures or uncertain BH mass estimates. This changed with the landmark discovery of EP240222a~\citep{Jin2025EP240222a}. First, the precise and prompt localization of the optical counterpart enabled timely spectroscopy, which both confirmed its redshift ($z = 0.03251 \pm 0.00013$) consistency with a nearby massive galaxy and obtained key TDE spectral features, including the characteristic $\mathrm{He\,\textsc{ii}}$ $\lambda 4686$\ and $\mathrm{N\,\textsc{iii}}$ $\lambda 4640$ lines. Second, both its X-ray spectra and host stellar mass indicate an IMBH with a mass most likely below $10^5$~\msun. EP240222a is thus the first robust IMBH-TDE captured in outburst, offering a promising path forward.

Interestingly, EP240222a exhibits a peculiar X-ray light curve: a multi-year slow rise, followed by a multi-month fast rise and a plateau lasting months to years~\citep{Jin2025EP240222a}. These features differ markedly from typical SMBH-TDEs and demand a thorough physical explanation. Establishing if the properties of this archetypal IMBH-TDE are universal or exceptional is critical for guiding future searches. Moreover, its deviation from SMBH-TDE light curves suggests a strong BH mass dependence, helping to pinpoint key parameters governing TDE evolution. In this Letter, we therefore focus on the multi-wavelength light curves of EP240222a. We propose a five-stage semi-phenomenological model to reproduce the observed behavior of these light curves and to illustrate the characteristic features of IMBH-TDEs.

In Section \ref{sec_obs}, we summarize the observational features of EP240222a, highlighting those that necessitate a new model. In Section \ref{sec_ana&mod}, we build our model for EP240222a upon the standard TDE framework, introducing key modifications. In Section \ref{sec_fit}, we apply this model to the X-ray light curve to constrain its parameters and interpret the physical implications of the fitting result. In Section \ref{sec_dis}, we present further discussion. In Section \ref{sec_con}, we summarize our conclusions.

We assume a cosmology with $H_0 = 70~\mr{km~s^{-1}~Mpc^{-1}}$, $\Omega_{\mr{m}} = 0.3$, and $\Omega_{\Lambda} = 0.7$.

\section{Observational Features} \label{sec_obs}

\begin{figure*} 
\centering
\includegraphics[width=\textwidth]{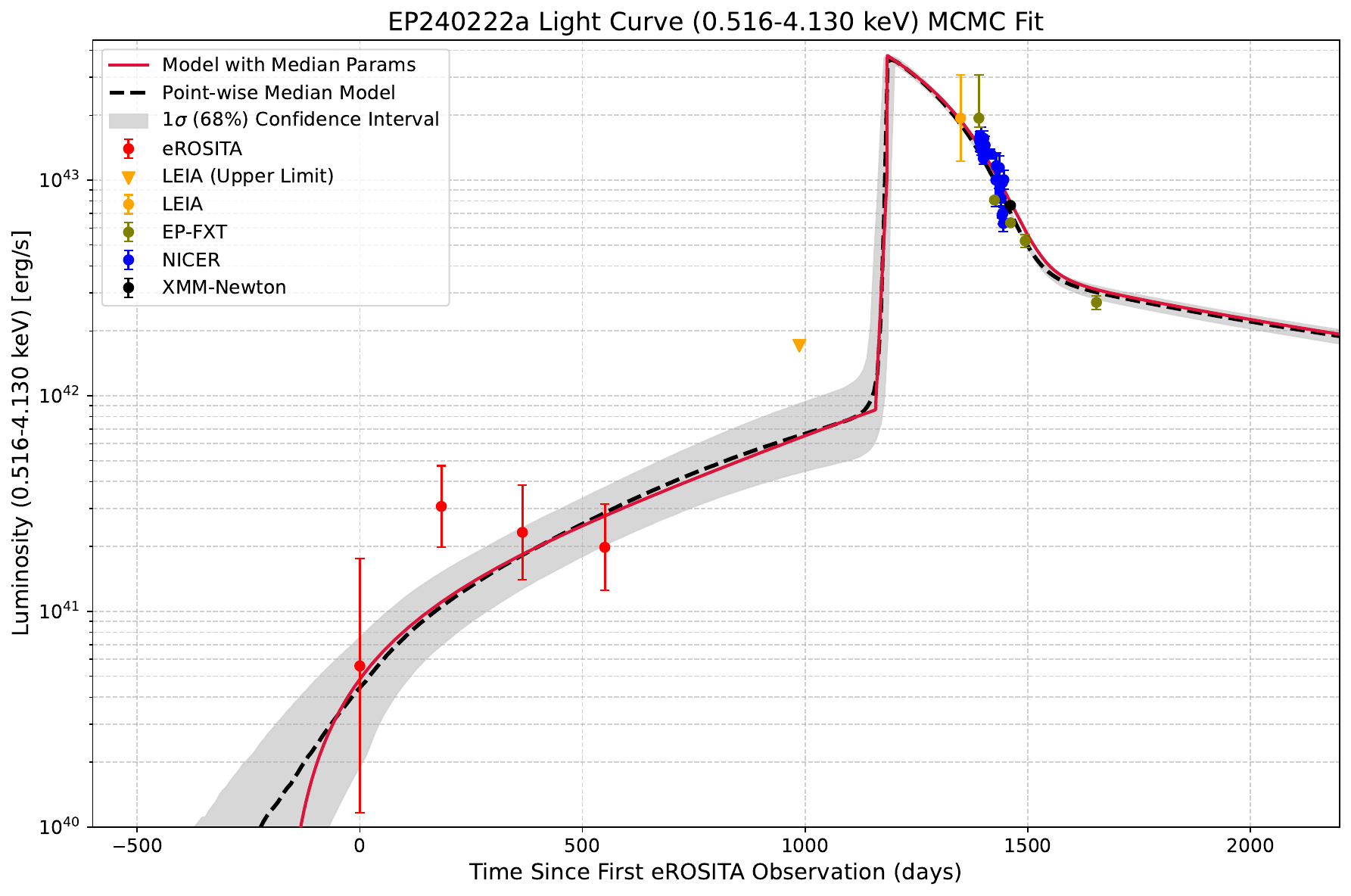} 
\caption{EP240222a X-ray light curve fit. Data: eROSITA (red), LEIA (orange), EP-FXT (olive), NICER (blue), XMM-Newton (black). The observed 0.5-4 keV band corresponds to 0.516–4.130 keV in the rest frame. Overplotted: median-parameter model (red solid), point-wise median model (black dashed), and the corresponding 1$\sigma$ (68\%) credible interval (gray shade). The model captures the distinct multi-stage evolution: a years-long slow rise, a sharp transition to the super-Eddington plateau (linear decay), and a subsequent power-law decline.} 
\label{fig:mcmc}
\end{figure*}

Before modeling the light curves of EP240222a, we first summarize its multi-wavelength observational features as reported in the discovery paper~\citep{Jin2025EP240222a}, emphasizing its unexpected characteristics.

\subsection{X-ray Discovery and Observational Features} \label{subsec_X_fea}

EP240222a was first discovered by the Wide-field X-ray Telescope (WXT) aboard the Einstein Probe (EP) satellite~\citep{Yuan2022,Yuan2025}. EP is an innovative soft X-ray satellite with a field-of-view of 3600 deg$^2$, providing unprecedented capability to characterize the dynamic X-ray sky. The discovery occurred on 2024 March 11, during EP's commissioning phase just two months after launch. Retrospective analysis revealed that the source was already bright two months prior, detected by both WXT and its pathfinder, LEIA~\citep{Zhang2022,Ling2023}, and had remained in a plateau stage during that period. It was later found that the source had initially been detected by eROSITA in 2020 with a slowly rising trend in following visits. \citet{Jin2025EP240222a} assume that the early eROSITA detections are physically linked to the later, more luminous X-ray emissions. This assumption is convincing, as both the early and late X-ray emissions have a similar spectral shape, which disfavours an AGN origin as it lacks a hard X-ray component. 

The X-ray properties of EP240222a exhibit some features consistent with theoretical predictions for IMBH-TDEs. For instance, its X-ray spectrum is relatively hard (inner disk temperature $\approx 200$~eV) compared to other non-jetted, thermal X-ray TDEs~\citep{Saxton2021,Grotova2025}. This is a natural consequence of the smaller black hole mass, which leads to a hotter inner accretion disk~\citep{Netzer2013}. Similarly, its relatively low peak luminosity of $L_{\mr{X,peak}} \sim 10^{43}$~\lum compared to other X-ray discovered TDEs~\citep{Sazonov2021,Grotova2025} is a direct result of its low-mass nature, with a smaller black hole having a lower Eddington limit~\citep{Netzer2013} that sets a lower ceiling for the peak luminosity~\citep{Chen2018}.

\subsection{Optical Features} \label{subsec_opt_fea}
As mentioned in the introduction, a key value of EP240222a lies in the timely detection of its optical counterpart during the X-ray plateau stage (see their Fig.~3; \citealt{Jin2025EP240222a}). However, the optical emission is remarkably faint ($L_\mr{g} \sim 10^{41}$~\lum), a factor of $\sim 200$ lower than the contemporaneous X-ray luminosity. Furthermore, archival Zwicky Transient Facility (ZTF) data reveal an optical plateau that began several months earlier. Unfortunately, the limited survey depth and observational gaps precluded any detections during the optical rise, yielding an upper limit of $L_{\mr{g,early}} \lesssim 10^{40}$~\lum. Following this plateau, the optical emission entered a decline stage that was both delayed and slower than the X-ray decline.

\subsection{Departures from IMBH-TDE Predictions}

The observations reveal several features that deviate significantly from previous predictions. First, the X-ray light curve (Fig.~\ref{fig:mcmc}) shows a remarkably slow rise over several years, evidenced by eROSITA detections. This was followed by a more rapid brightening of at least an order of magnitude in less than a year. Although unwitnessed, its timing is constrained between a LEIA upper limit in early 2023 and the subsequent LEIA detection in early 2024~\citep{Zhang2022, Ling2023}. This two-stage rise challenges models~\citep{Rees1988, Chen2018} that presume rapid circularization of fallback debris into a disk (see \S\ref{tdeprocess}).

Another unexpected feature is the faint optical and bright X-ray emissions during the peak plateau (see their Fig.~3; \citealt{Jin2025EP240222a}). This contrasts with the IMBH-TDE model of \citet{Chen2018} for a MS star. Their model predicts a years- to decade-long super-Eddington stage where X-rays are obscured by the ejected outflow. The observed emission is instead dominated by the outflow's reprocessed radiation, which peaks in the UV/optical. This persists until the accretion rate drops below Eddington, after which the bolometric luminosity decays as $t^{-5/3}$ and inner-disk X-rays emerge.


\section{Analysis and Modeling}  \label{sec_ana&mod}
\begin{figure*}
\gridline{
    \fig{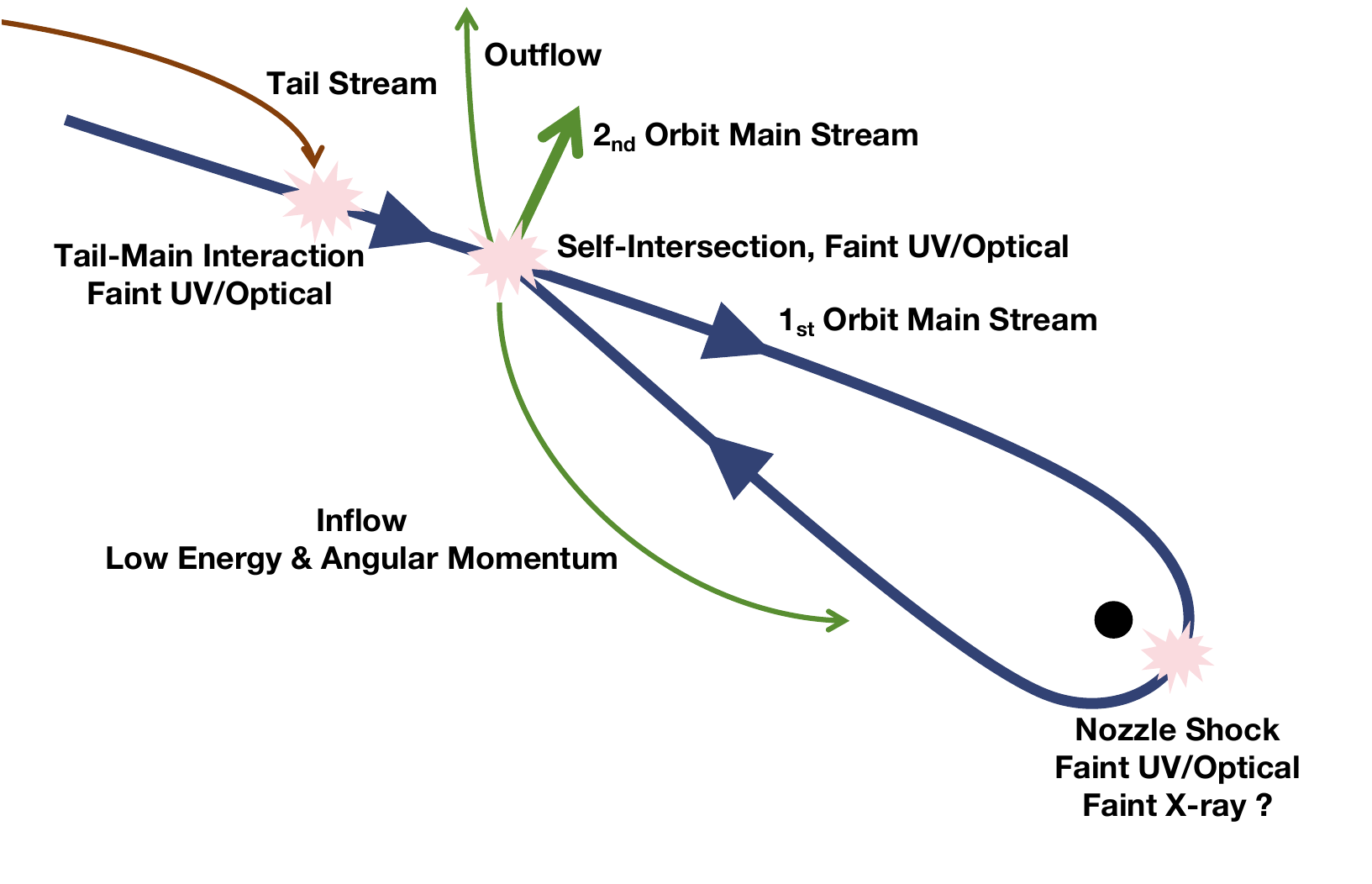}{0.5\textwidth}{\vspace{-10pt} (1) Initial Stage: Inefficient Circularization}
    \hspace{0pt}
    \fig{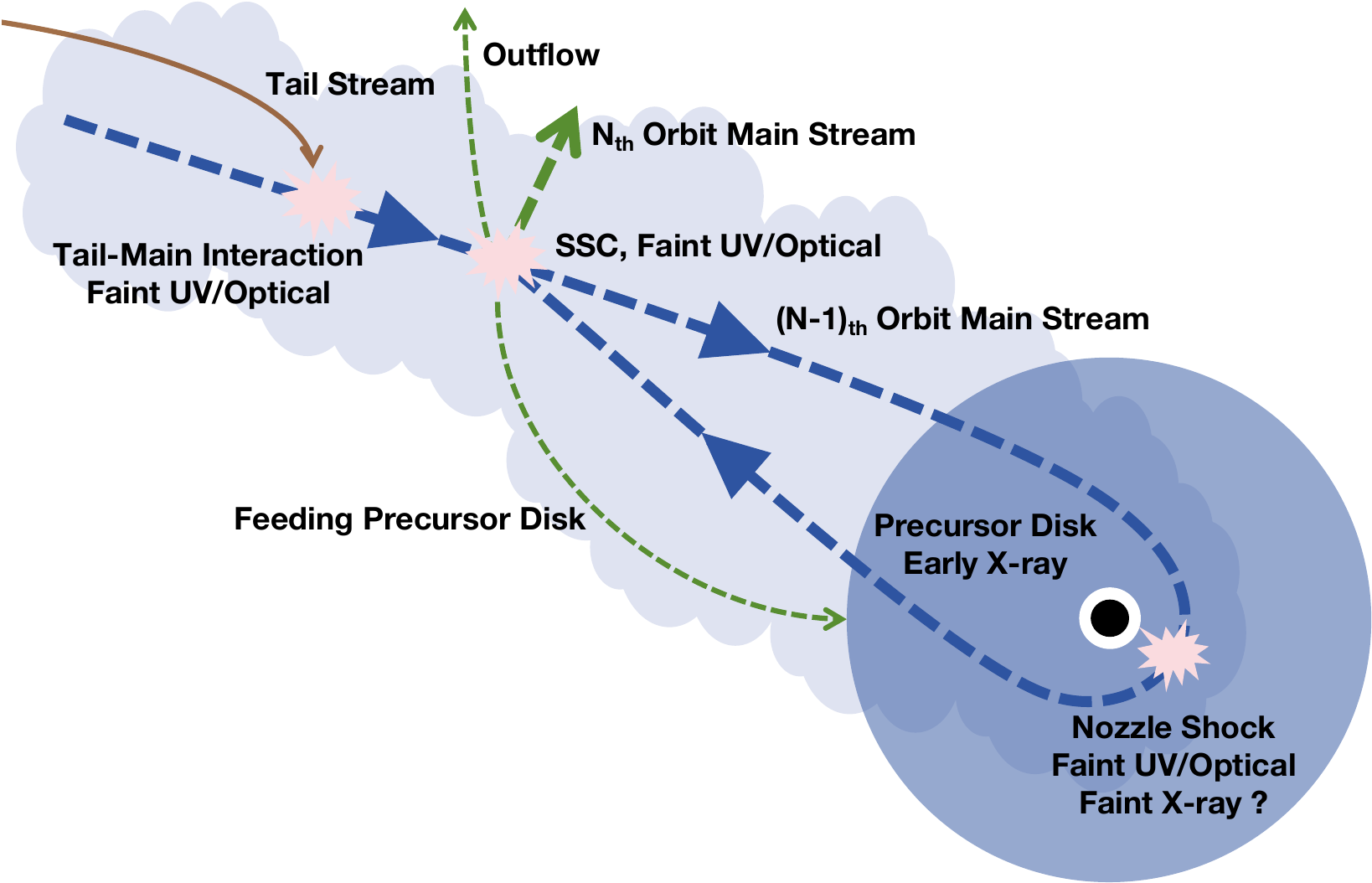}{0.5\textwidth}{\vspace{-10pt} (2) Slow-Rising Stage: SSC-Feeding Precursor Disk}
}
\vspace{-5pt}
\gridline{
    \fig{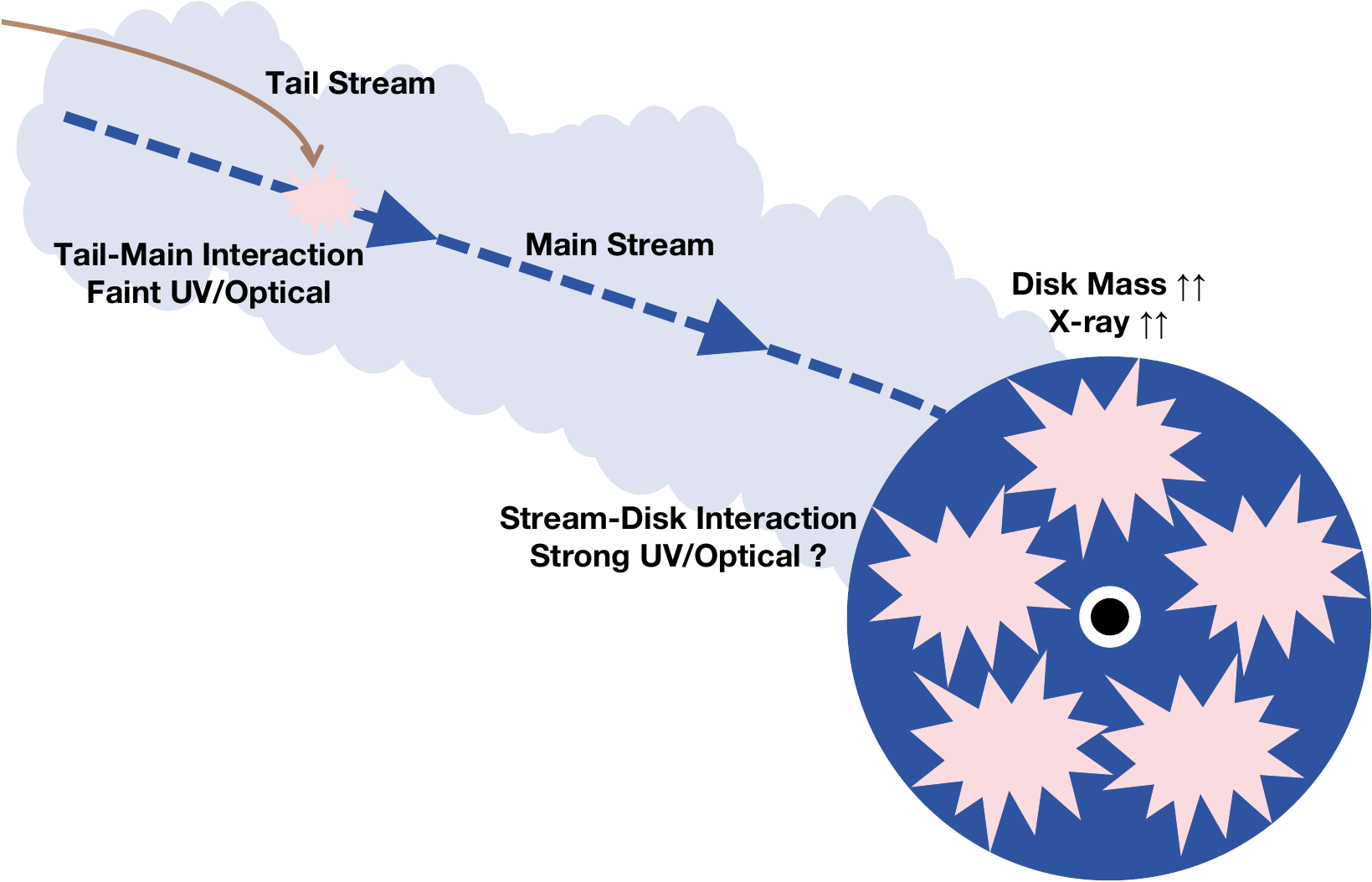}{0.5\textwidth}{\vspace{-10pt} (3) Fast-Rising Stage: Stream-Disk Interaction}
}
\vspace{-5pt}
\gridline{
    \hspace{0pt}
    \fig{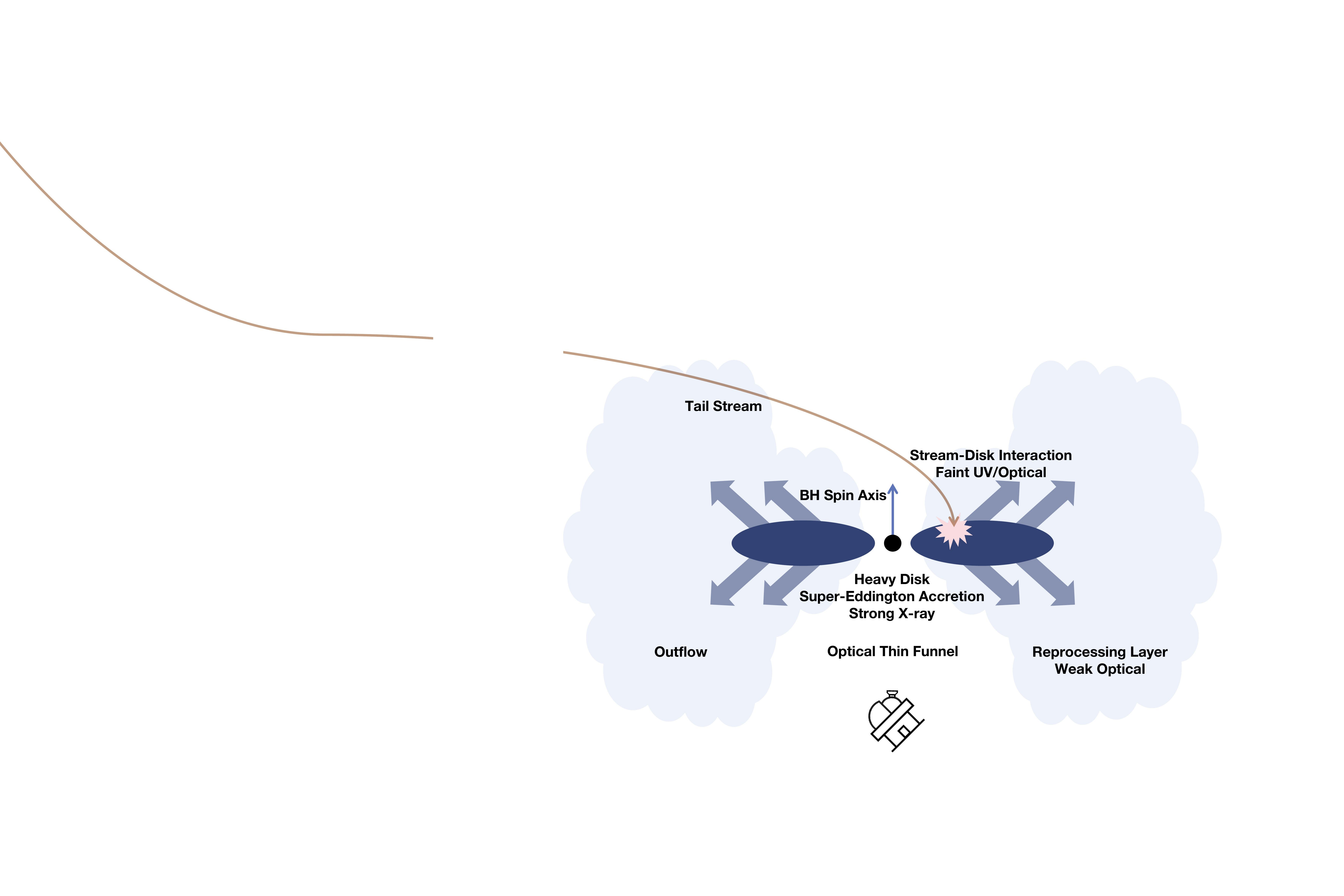}{0.46\textwidth}{\vspace{-10pt} (4) Plateau Stage: Super-Eddington Accretion}
    \hfill
    \hspace{0pt}
    \fig{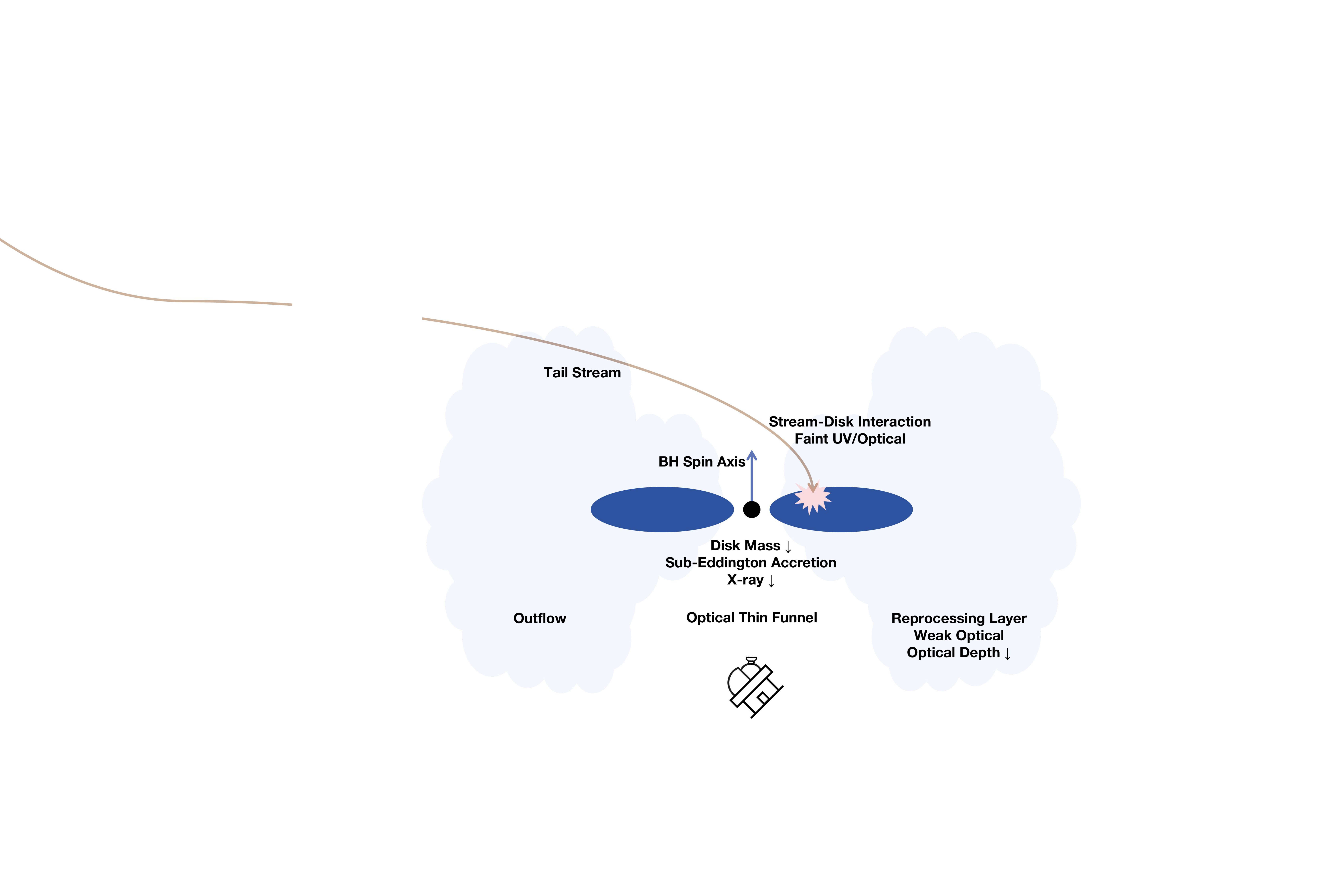}{0.46\textwidth}{\vspace{-10pt} (5) Decline Stage: Sub-Eddington Accretion}
}
\caption{
A schematic diagram illustrating the five-stage model. 
(1) \textbf{Initial Stage:} The three main inefficient dissipation channels (nozzle shock, self-intersection, tail-main stream interaction) produce faint emission. Self-intersection primarily drives both weak inflow and outflow. Timescale: $t_{\mr{fb}}$.
(2) \textbf{Slow-Rising Stage:} Within a few $t_{\mr{fb}}$, the main stream becomes thick from trapped radiation, while the tail remains thin. The dissipation processes continue (with the thick main stream's path shown schematically by the dashed lines), dominated by a succession of self-crossings (SSC). Inflow forms a precursor disk, which produces early X-rays. Low-luminosity X-rays lead to faint potential reprocessed UV/optical. Timescale: years.
(3) \textbf{Fast-Rising Stage:} Triggered by momentum flux matching between the disk and stream. A runaway circularization ensues, as the main stream violently merges into the disk. Disk mass builds up rapidly ($\sim t_{\mr{fb}}$), driving a sharp X-ray rise. Potential strong UV/optical flares might be suppressed by high optical depths. Torques from the main stream's merger rapidly align the disk with the BH's spin, thus misaligning it with the incoming tail stream. Timescale: $t_{\mr{fb}}$.
(4) \textbf{Plateau Stage:} A massive disk in a super-Eddington stage powers both outflow and Eddington-limited X-rays, which show a slow, linear decay. A polar funnel provides a clear line-of-sight to the X-rays. Weak optical arises from reprocessing by the outflow. The thin, dense tail stream penetrates the outflow to feed the disk, producing negligible light due to its low mass flux. Timescale: $t_{\mr{acc}}$.
(5) \textbf{Decline Stage:} As the disk mass depletes, the accretion rate drops to sub-Eddington. Consequently, X-rays transit from a linear to a power-law decay and the outflow ceases. The reprocessing layer expands and becomes optically thin over time (not yet reached). Timescale: years.
\textbf{Note:} Darker shades indicate higher density (for the thick stream in (2) and (3), density is represented by the dashed line's shade). Colors are for visual clarity only. Not to scale (reality: $R_{\mr{S}} \ll R_{\mr{p}} \sim R_{\mr{d}} \sim r_{\mr{s,p}} \ll a_0$).
}
\label{fig:four_stages_cartoon}
\end{figure*}

In this section, we develop a five-stage model to thoroughly understand the distinctive observational features of EP240222a. Our framework builds upon the standard TDE theory, retaining its successful components while introducing key modifications for IMBH-TDEs. A schematic diagram illustrating the model is shown in Fig.~\ref{fig:four_stages_cartoon}. For our analysis, we adopt the black hole mass $M_\mr{BH} = (7.7 \pm 4.0) \times 10^4~M_\odot$ and spin parameter $a_* = 0.98^{+0.02}_{-0.30}$ derived by \citet{Jin2025EP240222a} through spectral fitting with a slim disk model well-suited for (super-)Eddington accretion.

\subsection{A Brief Review of TDE Physical Process}
\label{tdeprocess}

The tidal radius $R_\mr{t} \equiv R_\mr{*}{(M_\mr{BH}/M_\mr{*})}^{1/3}$ marks the characteristic distance at which a star is disrupted by the BH's gravity, forming a debris stream at $t=0$. The encounter's intensity is quantified by the penetration factor $\beta \equiv R_{\mr T}/R_{\mr p}$, with $R_{\mr p}$ being the pericenter distance of the star's initial orbit. Following hydrodynamical simulations~\citep{Guillochon2013,Guillochon2015a}, the total mass of the bound debris $M_{\mr{fb}}$, which eventually falls back towards the BH, depends on $\beta$ and the star's internal structure parameterized by the polytropic index $\gamma$.

For a star with $\gamma=5/3$, the fallback mass is given by:
\begin{equation}
\label{eq:m_fb_5_3}
M_{\mr{fb}} = \frac{M_*}{2} \times \begin{cases} \exp\left[\frac{3.1647-6.3777\beta+3.1797\beta^2}{1-3.4137\beta+2.4616\beta^2}\right],&0.5 \lesssim \beta \lesssim 0.9 \\
1,&\beta \gtrsim 0.9.
\end{cases}
\end{equation}

For a star with $\gamma=4/3$, the relation is:
\begin{equation}
\label{eq:m_fb}
\begin{split}
M_{\mr{fb}} &= \frac{M_*}{2} \times \begin{cases} \exp\left[\frac{12.996-31.149\beta+12.865\beta^{2}}{1-5.3232\beta+6.4262\beta^{2}}\right] ,&0.6 \lesssim \beta \lesssim 1.85\\
1,&\beta \gtrsim 1.85.
\end{cases}
\end{split}
\end{equation}

The critical parameter $\beta_{\mr d}$ separates full tidal disruption events (FTDEs) and partial tidal disruption events (PTDEs). The threshold is $\beta_{\mr d} \approx 0.9$ for $\gamma=5/3$ and $\beta_{\mr d}\approx 1.85$ for $\gamma=4/3$.

Classical TDE theories~\citep{Rees1988} neglect both the circularization and accretion timescales ($t_{\mr{circ}}$ and $t_{\mr{acc}}$), assuming fallback material immediately forms a disk and is accreted. This implies the accretion rate directly follows the fallback rate, producing a light curve with a rapid rise ($t_\mr{rise} \sim 0.5~t_\mr{fb}$; \citealt{Evans1989}) followed by a power-law decline. Here, $t_{\mr{fb}}$ represents the orbital period of the most bound debris, given by:
\begin{equation}
\label{eq:t_fb}
\begin{split}
t_{\mr{fb}} &= 2 \pi \sqrt{\frac{a_0^3}{GM_{\mr{BH}}}} \\
&\simeq 13~ M_5^{1/2} r_*^{3/2} m_*^{-1}~ \text{days} \times \begin{cases}
\beta^{-3},&\beta \lesssim \beta_{\mr d} \\
1,&\beta \gtrsim \beta_{\mr d},
\end{cases}
\end{split}
\end{equation}
where $a_0$ is the semi-major axis of the most bound debris stream, $M_{\mr{BH}} \equiv M_5 \times 10^5 M_\odot$, $R_* \equiv r_* \times R_\odot$ and $M_* \equiv m_* \times M_\odot$. For typical parameters ($\beta=m_*=r_*=1$), $t_{\mr{fb}}$ is $\approx 4.1$, $13$, and $41$~days for a $10^4$, $10^5$, and $10^6 M_\odot$ BH, respectively.

However, this classical model fails for IMBH-TDEs such as EP240222a, as neither the circularization nor the accretion timescale is negligible, resulting in a long rise rather than a rapid one. We first demonstrate the inefficiency of primary energy dissipation mechanisms for prompt circularization in IMBH-TDEs of MS stars. Second, we model the long-term stream evolution and "precursor" disk formation in \S\ref{subsec_ssc}, which also accounts for the accretion timescale.

To quantify the energy dissipation rate, we model the debris stream as two components~\citep{Bonnerot2017,Chen2021,Chen2022}: a "main stream" representing the bulk of the returned material on a highly elliptical, precessing, slowly circularizing orbit, and a "tail stream" of continuously falling back material that supplies mass to the main stream. The tail's mass fallback rate follows a power law:
\begin{equation}
\label{eq:mdot_fb}
\dot{M}_{\mr{fb}}(t) = \dot{M}_{\mr{peak}} \left(\frac{t}{t_{\mr{fb}}}\right)^{-n},
\end{equation}
where the power-law index is $n=5/3$ for FTDEs and $n=9/4$~\citep{Coughlin2019,Miles2020} for PTDEs. The peak fallback rate is given by $\dot{M}_{\mr{peak}} = (n-1)M_{\mr{fb}}/t_{\mr{fb}}$. 

The mass of the main stream, $M_{\mr{s}}(t)$, is obtained by integrating the tail stream, assuming the tail stream is fully incorporated into the main stream, with any potential mass loss from the latter being negligible. This approximation, justified in \S\ref{subsec_main_stream_mass}, yields:
\begin{equation}
\label{eq:ms_t}
M_{\mr{s}}(t) = \int_{t_{\mr{fb}}}^{t} \dot{M}_{\mr{fb}}(t') dt'=\frac{\dot{M}_{\mr{peak}} t_{\mr{fb}}}{n-1} \left[ 1 - \left(\frac{t}{t_{\mr{fb}}}\right)^{1-n} \right].
\end{equation}

The specific energy required for circularization is the difference between the circularized and initial binding energies. The initial specific binding energy of the most bound debris is:
\begin{equation}
\label{eq:E_initial}
E_0 = \frac{GM_{\mr{BH}}}{2a_0} \simeq GM_{\mr{BH}}R_*\times
\begin{cases}
    R_\mr{p}^{-2},&\beta \lesssim \beta_{\mr{d}} \\
    R_\mr{T}^{-2},&\beta \gtrsim \beta_{\mr{d}}.
\end{cases}
\end{equation}

The most bound debris has an initial eccentricity $e_0 = 1-R_{\mr p}/a_0 \approx 1$. Assuming angular momentum conservation, the circularization radius is found to be $R_{\mr c} = R_{\mr p}(1+e_0) \simeq 2R_{\mr p}$. The final specific binding energy is therefore:
\begin{equation}
\label{eq:E_circ}
E_{\mr{c}} = \frac{G M_{\mr{BH}}}{2R_{\mr{c}}} \simeq \frac{G M_{\mr{BH}}}{4R_{\mr{p}}} \gg E_0.
\end{equation}

The total energy to be dissipated is then $\Delta E = E_{\mr{c}} - E_0 \simeq E_{\mr{c}}$.

\subsection{Initial Stage: Inefficient Circularization} \label{subsec_ini}

Three mechanisms drive early debris stream energy dissipation: nozzle shocks, main stream self-intersections, and tail-main stream interactions. Each involves supersonic collisions that generate shocks, dissipating kinetic energy from the relative motion. In this section, we analyze the efficiency of each process to demonstrate why initial circularization is highly inefficient for EP240222a.

We begin with the nozzle shock, which occurs near pericenter as the main stream is compressed by the BH's gravitational field. Its dissipation efficiency, defined as the energy dissipated per orbit relative to that needed for circularization, is~\citep{Bonnerot2021,Bonnerot2022}:
\begin{equation}
\label{eq:eff_nozzle}
\begin{split}
\epsilon_{\mr{noz}} &\simeq \frac{\frac{1}{2} v_{\mr{z,max}}^{\mr{2}}}{E_{\mr{c}}} \simeq 2 \times 10^{-2} ~ \beta \left(\frac{M_{\mr {BH}}}{M_*}\right)^{-2/3}
\\ &\simeq 9.3 \times 10^{-6} ~ \beta M_5^{-2/3} m_*^{2/3} \ll 1,
\end{split}
\end{equation}
where $v_{\mr{z,max}} \simeq 0.1 ~\beta \sqrt{GM_\mr{*}/{R_\mr{*}}}$ is the maximum vertical velocity component of the stream that is dissipated by the nozzle shock. For  $\beta=m_*=r_*=1$, $\epsilon_{\mr{noz}}$ is $\approx 4.3 \times 10^{-5}$, $9.3 \times 10^{-6}$, and $2.0 \times 10^{-6}$ for a $10^4$, $10^5$, and $10^6 M_\odot$ BH, respectively. However, this efficiency does not fully reduce the stream's orbital energy. The nozzle shock occurs in a highly optically thick medium~\citep{Bonnerot2022}, converting much of the dissipated energy into thermal and orbital energy rather than radiation. Consequently, the effective circularization efficiency is lowered as some energy is recycled into the orbit.

The second mechanism, main stream self-intersection due to General Relativistic (GR) apsidal precession, has a dissipation efficiency of~\citep{Dai2015,Bonnerot2021,Chen2021,Chen2022}:
\begin{equation}
\label{eq:eff_int}
\begin{split}
\epsilon_{\mr{self}} &\simeq \frac{\Delta E_\mr{0}}{E_{\mr{c}}} \simeq \frac{9\pi^2 G^2}{4c^4} e_{\mr{0}}^2 \beta^2 \frac{M_{\mr{BH}}^{4/3} M_*^{2/3}}{R_*^2}\\
             &\simeq 4.6 \times 10^{-4}~ e_{\mr{0}}^2 \beta^{2} M_5^{4/3} r_*^{-2} m_*^{2/3} \ll 1,
\end{split}
\end{equation}
where $\Delta E_\mr{0} = 9\pi^2 e_{\mr{0}}^2 G^3 M_{\mr{BH}}^3/(16c^4 {R_\mr{p}}^3)$ is the specific energy dissipated per orbit at the initial stage. For $e_0\approx\beta=m_*=r_*=1$, $\epsilon_{\mr{self}}$ is $\approx 2.2 \times 10^{-5}$, $4.6 \times 10^{-4}$, and $1.0 \times 10^{-2}$ for a $10^4$, $10^5$, and $10^6 M_\odot$ BH, respectively\footnote{The case of white dwarf disruptions, which may lead to more efficient circularization, is discussed in \S\ref{subsec_rise_timescale}}. However, like the nozzle shock, radiation trapping further reduces this efficiency. The self-intersection occurs near the main stream's apocenter, which is also highly optically thick at early times~\citep{Bonnerot2017, Bonnerot2021}, lowering the effective circularization efficiency.

The third mechanism, tail-main stream interaction, is dissipative but counterproductive to circularization. As discussed in \S\ref{subsec_main_stream_mass}, the tail stream merges smoothly with the main stream, primarily adding mass to the main stream and releasing energy. We define its efficiency as the energy released per fallback time relative to the circularization energy:
\begin{equation}
\label{eq:eff_tm}
\begin{split}
\epsilon_{\mr{tm}}(t) &\simeq \frac{\Delta E_\mr{tm}}{E_{\mr{c}}} \simeq \frac{\dot{M_\mr{fb}}(t) t_\mr{fb} E_0}{M_\mr{s}(t) E_\mr{c}} \\
             &\simeq 4.3 \times 10^{-2}~ (\frac{t}{t_\mr{fb}})^{-n} \left[ 1 - \left(\frac{t}{t_{\mr{fb}}}\right)^{1-n} \right]^{-1} (n-1) (1+e_0)  \\
            & \quad \times M_5^{-1/3} m_*^{1/3}\times\begin{cases}
    \beta, & \beta \lesssim \beta_{\mr{d}} \\
    \beta^{-1}, & \beta \gtrsim \beta_{\mr{d}}
\end{cases} \\
    &\ll 1,
\end{split}
\end{equation}
where we approximate that the main stream's specific binding energy remains at its initial value and the tail stream's is negligible. This applies only for $t \gtrsim 2.5~t_{\mr{fb}}$, as earlier interactions constitute the main stream's first self-intersection. Early tail-main stream interaction occurs near apocenter, where, like self-intersection, most energy is trapped as internal and kinetic energy instead of being radiated. This makes the interaction an anti-circularization process that slightly increases the main stream's specific energy.

In summary, for IMBHs disrupting MS stars, nozzle and self-intersection shocks are inefficient, while tail-main interaction impedes circularization. The system thus "stalls" after the first few orbits: the main stream remains highly eccentric ($e\simeq e_0 \approx 1$) and minimal binding energy ($E\simeq E_0 \ll E_{\mr{c}}$), preventing rapid circularization and necessitating evolution over longer timescales. Simultaneously, self-intersection drives inflow and outflow (detailed in \S\ref{subsec_main_stream_mass}), with the inflow gradually forming the crucial precursor disk modeled in \S\ref{subsec_ssc}.

\subsection{Slow-Rising Stage: SSC-Feeding Precursor Disk} \label{subsec_ssc}

Having established the initial inefficiency of circularization, we now examine how the dissipation channels evolve to explain the slow X-ray rise. The relative importance of the three mechanisms (see \S\ref{subsec_ini}) shifts significantly after a few orbits. The key evolution is the main stream's geometric thickening. Initially, high optical depth traps radiation, driving expansion over several~$t_{\mr{fb}}$ until the diffusion timescale $t_{\mr{diff}}$ drops near the orbital period, allowing efficient photon escape~\citep{Bonnerot2017, Bonnerot2021}. This forms a thick, highly eccentric stream filling a volume extending to $\sim a_0$\footnote{The actual expansion radius can be slightly larger than $a_0$ for the BH mass of EP240222a (see Eq.~18 of~\citealp{Bonnerot2021}). However, it remains on the order of a few $a_0$, and any resulting difference in the energy dissipation rate is absorbed into our phenomenological efficiency factor, $\eta_{\mr{self}}(t)$.}. Despite this expansion, angular momentum conservation maintains a well-defined pericenter at $\sim R_{\mr{p}}$, with confined spatial dispersion near this point (stream cross section $r_\mr{s,p} \sim R_\mr{p}$). Meanwhile, the tail stream remains comparatively thin.

Stream thickening dramatically alters the dissipation mechanisms. The nozzle shock weakens or ceases as the thick stream avoids strong pericenter compression~\citep{Bonnerot2022}. Though tail-main stream interaction persists (Eq.~\ref{eq:eff_tm}), its strength declines with the fallback rate, becoming negligible after tens of $t_{\mr{fb}}$. However, main stream self-intersection remains a viable dissipation channel. Simulations~\citep{Bonnerot2020,Bonnerot2021} show that while more chaotic, the process persists over a broader range of radii, dominating long-term energy dissipation.

Therefore, we model the long-term evolution using the succession of self-crossings (SSC) framework~\citep{Dai2015,Bonnerot2021,Chen2021,Chen2022}. Within this framework, the main stream progressively loses orbital energy via successive self-intersections, causing it to gradually spiral inward. The idealized specific energy dissipation rate, $\dot{E}_{\mr{ideal}}(t)$, assuming a thin main stream and perfect collisions, is given by~\citep{Chen2021}:
\begin{equation}		
\label{eq:diss}
\dot{E}_{\mr{ideal}}(t) \simeq \frac{\Delta E_0}{t_{\mr{fb}}} \frac{1}{e_0^2} \left(1 - \frac{E_{\mr{ideal}}(t)}{E_{\mr{c}}}\right) \left(\frac{E_{\mr{ideal}}(t)}{E_0}\right)^{3/2}.
\end{equation}
We initiate this dissipation at the first self-intersection ($t \simeq 1.5~t_{\mr{fb}}$), with zero dissipation prior to this. The initial conditions are thus $E_{\mr{ideal}}(1.5~t_\mr{fb}) = E_0$ and $e_{\mr{ideal}}(1.5~t_\mr{fb}) = e_0$.

To account for any deviations from the idealized model, we further introduce an efficiency factor, $\eta_\mr{self}(t)$. The actual energy dissipation rate, $\dot{E}(t)$, is then modeled as:
\begin{equation}
\label{eq:diss_actual}
\dot{E}(t) = \eta_{\mr{self}}(t) \dot{E}_{\mr{ideal}}(t).
\end{equation}
While the idealized rate $\dot{E}_{\mr{ideal}}(t)$ increases rapidly, the actual rate $\dot{E}(t)$ is expected to rise more slowly. This is due to the likely small $\eta_\mr{self}(t)$ constrained in \S\ref{sub_sub_sec_early_opt} via the faint early optical emission, keeping $E(t)$ close to $E_0$ and thus suppressing the growth of $\dot{E}(t)$ (see Eq.~\ref{eq:diss}). Therefore, $\eta_{\mr{self}}(t)$ should be a decreasing function of time.

The main stream emits X-rays efficiently only after circularizing into a disk. However, simulations~\citep{Shiokawa2015,Bonnerot2020,Andalman2022} show that self-intersection shocks can cause a small fraction of material to lose excess energy and angular momentum, fall near the BH, undergo complex interactions, and then form a low-mass precursor disk that may explain faint early X-ray emission in some TDEs~\citep{Liu2022,Hajela2025}. Although this material has low energy and angular momentum relative to the main stream, the change is of order unity; thus, for simplicity, we assume the precursor disk forms at a radius of $R_\mr{c}$.

The mass supply to the precursor disk is complex and subject to perturbations from the main stream. We model it as proportional to the main stream's energy dissipation rate\footnote{Contributions to the mass supply from nozzle shocks and tail-main stream interaction are negligible compared to self-intersection (see \S\ref{subsec_main_stream_mass}).}:
\begin{equation}
\label{eq:mdot_sup}
\begin{split}
\dot{M}_{\mr{sup}}(t) 
& \simeq \eta_{\mr{sup}}(t) M_{\mr{s}}(t) \frac{\dot{E}(t)}{E_{\mr{c}}-E_0} \simeq \eta_{\mr{self}}(t) \eta_{\mr{sup}}(t) M_{\mr{s}}(t) \frac{\dot{E}_\mr{ideal}(t)}{E_{\mr{c}}-E_0} \\
& \simeq \overline{\eta_{\mr{self}} \eta_{\mr{sup}}} M_{\mr{s}}(t) \frac{\dot{E}_\mr{ideal}(t)}{E_{\mr{c}}-E_0}
\end{split}
\end{equation}
where the mass-supply efficiency, $\eta_{\mr{sup}}(t)$, expected to increase over time as the main stream becomes more chaotic. The final equality introduces our first-order, long-term averaged, semi-phenomenological approximation: we assume the product of the decreasing $\eta_{\mr{self}}(t)$ and increasing $\eta_{\mr{sup}}(t)$ is a slowly varying quantity, approximated by the constant parameter $\overline{\eta_{\mr{self}} \eta_{\mr{sup}}}$.

Next, we evaluate the accretion timescale $t_{\mr{acc}}$ by adopting the standard $\alpha$-disk model:
\begin{equation}
\label{eq:t_acc}
\begin{split}
t_{\mr{acc}} & \simeq \left(\frac{H}{R}\right)_\mr{d}^{-2} \alpha^{-1} t_{\mr{dyn,c}} \\
             & \simeq 13~ \left(\frac{(H/R)_\mr{d}}{0.2}\right)^{-2}
             \left(\frac{\alpha}{0.1}\right)^{-1} \beta^{-3/2} r_*^{3/2} m_*^{-1/2}~ \text{days},
\end{split}
\end{equation}
where $(H/R)_\mr{d}$ is the disk aspect ratio, $\alpha$ the viscosity parameter, and $t_{\mr{dyn},c} \equiv \sqrt{R_{\mr c}^3/(G M_{\mr {BH}})}$ the dynamical timescale at the circularization radius, $R_{\mr c}$. Using typical values from simulations, such as $(H/R)_\mr{d} \approx 0.1-2$~\citep{Sadowski2016,Bonnerot2020,Andalman2022,Steinberg2024} and $\alpha \approx 0.02 - 0.4$~\citep{Sadowski2016,Bonnerot2020}, and assuming $\beta=m_*=r_*=1$, yields $t_{\mr{acc}} \simeq (0.625 - 5000)~t_{\mr{dyn,c}} \approx (0.03 - 261)$~days. This implies $t_{\mr{acc}}$ can be comparable to, or even significantly longer than $t_{\mr{fb}}$, and therefore cannot be neglected either. 

For the precursor disk, chaotic conditions likely hinder magnetorotational instability (MRI) development, implying inefficient angular momentum transport and thus smaller viscosity parameter, $\alpha_\mr{early}$. Additionally, the lower accretion rate suggests a smaller aspect ratio, $(H/R)_\mr{d,early}$, likely intermediate between a classical, gas-pressure-dominated thin disk ($(H/R)_\mr{d} \sim 0.001-0.01$; \citealt{Shen2014}), and the much thicker disks seen in TDE simulations. These combined effects push $t_{\mr{acc}}$ towards the upper estimates, reinforcing its non-negligibility.

The mass continuity equation, used throughout to track disk evolution, is:
\begin{equation}
\label{eq:mass_continuity}
\frac{dM_{\mr{d}}(t)}{dt} = \dot{M}_{\mr{sup}}(t) - \dot{M}_{\mr{acc}}(t),
\end{equation}
with accretion rate  $\dot{M}_{\mr{acc}}(t) = M_{\mr{d}}(t) / t_{\mr{acc,early}}$.

It is important to note that we adopt distinct timescales: $t_{\mr{acc, early}}$ (initial/slow-rising) and $t_{\mr{acc,late}}$ (subsequent stages), reflecting expected differences in viscosity and disk structure between these two distinct phases as argued above.

The bolometric luminosity during the early rising stage is:
\begin{equation}
\label{eq:l_bol}
L_{\mr{bol}}(t) \simeq \eta_{\mr{mis}} \dot{M}_{\mr{acc}}(t) c^2,
\end{equation}
where $\eta_{\mr{mis}}$ is the radiative efficiency for a potentially misaligned disk, bounded by retrograde/prograde limits:
\begin{equation}
\label{eq:eta_bounds}
\eta_{\mr{retro}} < \eta_{\mr{mis}} < \eta_{\mr{align}}.
\end{equation}
For the BH here ($a_* \approx 0.98$), these theoretical bounds are $\eta_{\mr{retro}} \approx 3.8\%$ and $\eta_{\mr{align}} \approx 23.4\%$~\citep{Bardeen1972}.

\subsection{Fast-Rising Stage: Stream-disk Interaction} \label{subsec_rising}
Although unwitnessed, a rapid brightening stage is inferred from the order-of-magnitude jump between eROSITA's slow rise and the 2024 plateau (Fig.~\ref{fig:mcmc}; Fig.~3 of \citealt{Jin2025EP240222a}). This implies that a physical mechanism, initially suppressed, was triggered once the precursor disk reached a critical mass, consistent with recent simulations~\citep{Andalman2022,Steinberg2024}, where debris stream interactions with a precursor disk trigger rapid circularization.

Initially ($\sim$ several $t_{\mr{fb}}$ post-disruption), the massive main stream ($M_\mr{s}(t) \simeq M_{\mr{fb}}$) can penetrate the low-mass precursor disk largely unimpeded. As slow circularization continues, the disk's mass and density grow. A critical point is reached at $t_{\mr{crit}}$ when the effective momentum flux of the main stream matches that of the disk.

This hinders penetration, triggering a violent runaway merger: the interaction increases disk mass and density, thereby enhancing interception and accelerating the process. Lasting for the orbital timescale of the $N$-th orbit ($t_\mr{s,N} = \pi G M_\mr{BH}/\sqrt{2{E(t_\mr{crit})}^3}\simeq t_\mr{fb}$), this process drives full main stream assimilation, efficiently dissipating orbital energy and leading to rapid circularization. The physical justification for this model is further explored in \S\ref{subsec_stream_disk} in light of our fitting results.

We model mass supply as the sum of rapid main stream incorporation (dominant; over one orbit) and continuous tail fallback (subdominant):
\begin{equation}
\label{eq:mdot_sup_2}
\dot{M}_{\mr{sup}}(t) \simeq \frac{M_{\mr s}(t)}{t_{\mr{s,N}}} + \dot{M_\mr{fb}}(t).
\end{equation}

We neglect direct radiation from this violent interaction due to high optical depths (likely converting a large fraction of the energy into kinetic forms) and the lack of data on such relatively short timescales. Furthermore, the short duration $t_{\mr{s,N}}$ implies negligible impact on the subsequent, longer-timescale plateau and decline stages we aim to model.

We therefore focus on the outcome of this interaction: the rapid mass supply to the disk, boosting $\dot{M}_{\mr{acc}}(t)$ to super-Eddington within $t_\mr{s,N}$, and driving the rapid X-ray rise. Meanwhile, outflow is launched from the disk, reducing the actual mass accretion rate onto the event horizon of the BH to $\sim \dot{M}_\mr{edd}$ \footnote{$\dot{M}_{\mr{acc}}(t) = M_{\mr{d}}(t) / t_{\mr{acc}}$ still holds, because $\dot{M}_{\mr{acc}}(t)$ represents the accretion rate in the outer regions of the disk, before significant mass loss by outflow.}, effectively "throttling" radiative output. This saturates luminosity at $\simeq L_{\mr{Edd}}$ despite $\dot{M}_{\mr{acc}}(t) \gg \dot{M}_{\mr{Edd}}$.

The luminosity follows\footnote{We neglect order-unity corrections to the logarithmic term arising from advection during super-Eddington accretion~\citep{Chashkina2019}, outflow reprocessing, and deviations where $R_\mr{c}< \text{spherization radius}$ for simplicity.}~\citep{King2016}:
\begin{equation}
\label{eq:L_edd}
L_{\mr{bol}}(t) \simeq
\begin{cases}
    \left[1 + \ln\left(\dot{M}_{\mr{acc}}(t)/\dot{M}_{\mr{Edd}}\right)\right]L_{\mr{Edd}}, & \dot{M}_{\mr{acc}}(t) \gtrsim \dot{M}_{\mr{Edd}}  \\
    \left(\dot{M}_{\mr{acc}}(t) / \dot{M}_{\mr{Edd}}\right)L_{\mr{Edd}}, & \dot{M}_{\mr{acc}}(t) \lesssim \dot{M}_{\mr{Edd}}.
\end{cases}
\end{equation}
where the Eddington luminosity is $L_{\mr{Edd}} \simeq 1.26 \times 10^{43}~M_5~\mr{erg~s^{-1}}$, and the corresponding Eddington accretion rate, $\dot{M}_{\mr{Edd}}$, is calculated via $L_{\mr{Edd}} = \eta_{\mr{align}}\dot{M}_{\mr{Edd}}c^{2}$. We adopt $\eta_{\mr{align}} \approx 23.4\%$ as powerful stream-disk merger torques during the fast-rising stage drive rapid alignment of the disk with the BH spin~\citep{Zanazzi2019}. Consequently, the disk becomes misaligned with the subsequent infalling tail.

\subsection{Plateau Stage: Super-Eddington Accretion} \label{subsec_plateau}
During this stage, the disk, massive from rapid main stream injection, sustains prolonged super-Eddington accretion, saturating luminosity near $L_{\mr{Edd}}$ (Eq.~\ref{eq:L_edd}). The plateau is not perfectly flat but shows a slow, linear decline (a "ln(A-x)"-like decay in semi-log space; see Fig.~\ref{fig:mcmc}).

This can be understood as follows. After the fast rise, with a large mass $\simeq M_{\mr{fb}}$ and negligible tail supply, Eq.~\ref{eq:mass_continuity} simplifies to $dM_{\mr{d}}(t)/{dt} \simeq -\dot{M}_{\mr{acc}}(t) = -M_{\mr{d}}(t)/t_{\mr{acc,late}}$. Substituting the solution $\dot{M}_\mr{acc}(t) \simeq M_{\mr{fb}} \exp(-t/t_{\mr{acc,late}})/t_\mr{acc,late}$ into Eq.~\ref{eq:L_edd}:
\begin{equation}
\label{eq:L_log_linear_decay}
L_{\mr{bol}}(t) \simeq \left[ 1+\ln\left(\frac{M_{\mr{fb}}}{\dot{M}_{\mr{Edd}}t_{\mr{acc,late}}}\right) - \frac{t}{t_{\mr{acc,late}}} \right] L_\mr{Edd}.
\end{equation}
This predicts a linear bolometric luminosity decline (appearing as "ln(A-x)"-like decay in semi-log space), providing a precise physical description: the visually identified "plateau" corresponds to the early, brighter, flat portion of this decay, while the initial phase of the "decline" stage corresponds to its subsequent, steeper portion.

Supported by observations (see \S\ref{Plateau_and_Decline} and \citealp{Jin2025EP240222a}), continuous outflows form a reprocessing layer~\citep{Strubbe2009,Chen2018} converting a fraction of the central X-rays to longer wavelengths, naturally explaining the optical emergence and plateau. A Polar funnel~\citep{Dai2018, Qiao2025} provides a clear sightline, revealing the unobscured Eddington-limited central X-rays.

Despite the outflow, the dense, thin tail penetrates to feed the disk directly at $\dot{M}_\mr{fb}(t)$, though associated interaction luminosity is negligible due to low mass flux.

\subsection{Decline Stage: Sub-Eddington Accretion} \label{subsec_decline}
The decline initially follows the steeper "ln(A-x)"-like decay. After $\sim t_{\mr{acc,late}}$, reservoir depletion drops accretion below Eddington. With the reservoir exhausted, accretion tracks the continuous tail fallback ($\dot{M}_{\mr{acc}}(t) \simeq \dot{M}_{\mr{fb}}(t)$; luminosity following Eq.~\ref{eq:L_edd}). Consequently, the light curve transitions to a shallower, fallback-tracking power-law decay (linear in semi-log space). The final two X-ray points hint at this flattening. We caution, however, late-time evolution may involve complex structural and temperature changes as accretion drops further.


\section{X-ray Light Curve Fitting} \label{sec_fit}

\subsection{Parameter Setup}
We constrain seven free parameters via Markov Chain Monte Carlo (MCMC) fitting to the X-ray light curve, adopting flat priors over physically motivated ranges:
\begin{enumerate}
    \item $\textbf{Stellar Mass, } \log_{10}(M_*/M_\odot) \in [-1.097, 0.477]$ (i.e., $0.08-3~M_\odot$). This spans the MS range~\citep{Kippenhahn1994}, excluding rare massive~\citep{Kochanek2016} and non-MS stars.
    
    \item $\textbf{Penetration Factor, } \log_{10}(\beta) \in [-0.301, 0.477]$ (i.e., $\beta \in [0.5, 3]$). This spans the disruption threshold to rare deep encounters~\citep{Coughlin2022}.

    \item $\textbf{Efficiency Factor, } \log_{10}(\eta) \equiv \log_{10}(\overline{\eta_{\mr{self}} \eta_{\mr{sup}}} \eta_{\mr{mis}}) \in [-6, 1]$. Upper bound exceeding 0 allows chaotic self-intersections to enhance $\eta_{\mr{self}}(t)$.

    \item $\textbf{Time Offset, } t_0 \in [0, 1000]$ days, representing the interval between disruption and the first eROSITA data.

    \item $\textbf{Early Accretion Timescale, } \log_{10}(t_{\mr{acc, early}}/t_{\mr{dyn,c}}) \in [0, 9]$. The lower bound enforces $t_{\mr{acc}} \ge t_{\mr{dyn}}$, while the large upper bound accommodates very low viscosity and/or aspect-ratio precursor disks (see \S\ref{subsec_ssc}).

    \item $\textbf{Late Accretion Timescale, } \log_{10}(t_{\mr{acc, late}}/t_{\mr{dyn,c}}) \in [0, 4]$, covering typical values with margin (\S\ref{subsec_ssc}). Note that allowing $t_{\mr{acc,late}} > t_{\mr{acc,early}}$ can produce a light curve dip between the slow- and fast-rising stages.

    \item $\textbf{Critical Time, } t_{\mr{crit}} - t_0 \in [950, 1300]$ days. This marks the transition to the fast-rising stage, relative to the first eROSITA detection, constrained by visual inspection. Physical plausibility of the transition is discussed in \S\ref{subsec_stream_disk}.

\end{enumerate}

Stellar radius $R_*$ is derived from $M_*$ via the MS mass-radius relation~\citep{Kippenhahn1994}:
\begin{equation}
\label{eq:mass_radius}
\frac{R_*}{R_\odot} = \begin{cases}
(\frac{M_*}{M_\odot})^{0.80},&\quad{0.08 \lesssim \frac{M_*}{M_\odot} \lesssim 1} \\
(\frac{M_*}{M_\odot})^{0.57},&\quad{1 \lesssim \frac{M_*}{M_\odot} \lesssim 3}.
\end{cases}
\end{equation}

Similarly, the polytropic index $\gamma$ (describing the star's internal structure) depends on $M_*$~\citep{Mockler2019}:
\begin{equation}
\label{eq:gamma_mass}
\gamma = \begin{cases}
5/3,&\quad{0.08 \lesssim \frac{M_*}{M_\odot} \lesssim 0.3} \\
\text{Linear Interpolated},&\quad{0.3 \lesssim \frac{M_*}{M_\odot} \lesssim 1} \\
4/3,&\quad{1 \lesssim \frac{M_*}{M_\odot} \lesssim 3}.
\end{cases}
\end{equation}
In the transition region ($0.3 \lesssim \frac{M_*}{M_\odot} \lesssim 1$), rather than interpolating $\gamma$, we linearly interpolate a minimal set of fundamental physical quantities between pure $\gamma=5/3$ and $\gamma=4/3$ models. These interpolated fundamentals are then used to calculate all derived quantities, minimizing interpolation artifacts.

\subsection{Bolometric Correction}
To compare model $L_{\mr{bol}}$ with observed 0.5-4~keV unabsorbed fluxes~\citep{Jin2025EP240222a}, we apply a bolometric correction (BC). Given the minimal temperature evolution of EP240222a (typical of TDEs), we adopt a constant BC.

Calculating BCs for two epochs with different methods yields consistent results: combined eRASS (blackbody, $T_\mr{X} = 158$~eV) gives $K_{\mr{X}} \approx 1.82$, while EP-FXT epoch-1 ({\tt{TBabs*tdediscspec}}) gives $K_{\mr{X}} \approx 2.24$. We adopt the latter, as it is physically more representative of a TDE disk. 

Significant unobserved emission ("missing energy"; \citealt{Dai2018,Guolo2024,Qiao2025}) implies larger BCs but is unlikely here. First, X-ray spectra peaking at a few hundred eV capture the spectral energy distribution (SED) peak, suggesting bulk emission is observed. Second, the optical luminosity is $\sim 200$ times fainter than X-rays during the plateau stage. If a large amount of energy were being emitted in an intermediate, unobserved band, we would expect the reprocessed optical luminosity to be comparable to the X-rays. This contradicts observations, thus confirming that total power is overwhelmingly dominated by X-rays (see \S\ref{subsec_opt_light_curve} for detailed discussion of the weak optical).

\subsection{MCMC Fitting Procedure}

Fitting uses the MCMC package {\tt{emcee}}~\citep{Daniel2013}, utilizing the {\tt{emcee.moves.DEMove}} with an ensemble of 94 walkers, each taking $400,000$ steps.

Assuming Gaussian errors, the log-likelihood $\ln \mathcal{L}$ sums detected ($N_{\mr{det}}$) and upper-limit ($N_{\mr{ul}}$) components. For detections, the log-likelihood is the  standard chi-squared form:
\begin{equation}
\label{eq:loglike_det}
\begin{split}
\ln \mathcal{L}_{\mr{det}} = -\frac{1}{2} \sum_{i=1}^{N_{\mr{det}}}\begin{cases}
\left( \frac{ L_{\mr{obs}, i} - L_{\mr{model}}(t_i)}{\sigma_{i,+}} \right)^2,&\quad{L_{\mr{obs}, i} \leq L_{\mr{model}}} \\
\left( \frac{ L_{\mr{obs}, i} - L_{\mr{model}}(t_i)}{\sigma_{i,-}} \right)^2,&\quad{L_{\mr{obs}, i} > L_{\mr{model}}},
\end{cases}\\
\end{split}
\end{equation}
where $L_{\mr{model}}(t_i)$ is the predicted luminosity and $\sigma_{i,\pm}$ denote upper/lower error bounds on the observed luminosity $L_{\mr{obs}, i}$.

For $N_{\mr{ul}}$ upper limits (here $N_{\mr{ul}}=1$), the log-likelihood is:
\begin{equation}
\label{eq:loglike_ul}
\ln \mathcal{L}_{\mr{ul}} = \sum_{j=1}^{N_{\mr{ul}}} \ln \left[ \frac{1}{2} \left( 1 + \operatorname{erf} \left( \frac{L_{\mr{ul}, j} - L_{\mr{model}}(t_j)}{\sqrt{2}\sigma_j} \right) \right) \right],
\end{equation}
where $\operatorname{erf}$ is the error function. For 90\% confidence level (CL) upper limits $L_{\mr{ul}, j}$, the effective standard deviation $\sigma_j = L_{\mr{ul}, j} / Z$, with $Z \approx 1.28$ the Z-score corresponding to this CL. 

The total log-likelihood is then $\ln \mathcal{L} = \ln \mathcal{L}_{\mr{det}} + \ln \mathcal{L}_{\mr{ul}}$.

\subsection{Fitting Results} \label{subsec_fitting_results}

\begin{figure*} 
\centering
\includegraphics[width=\textwidth]{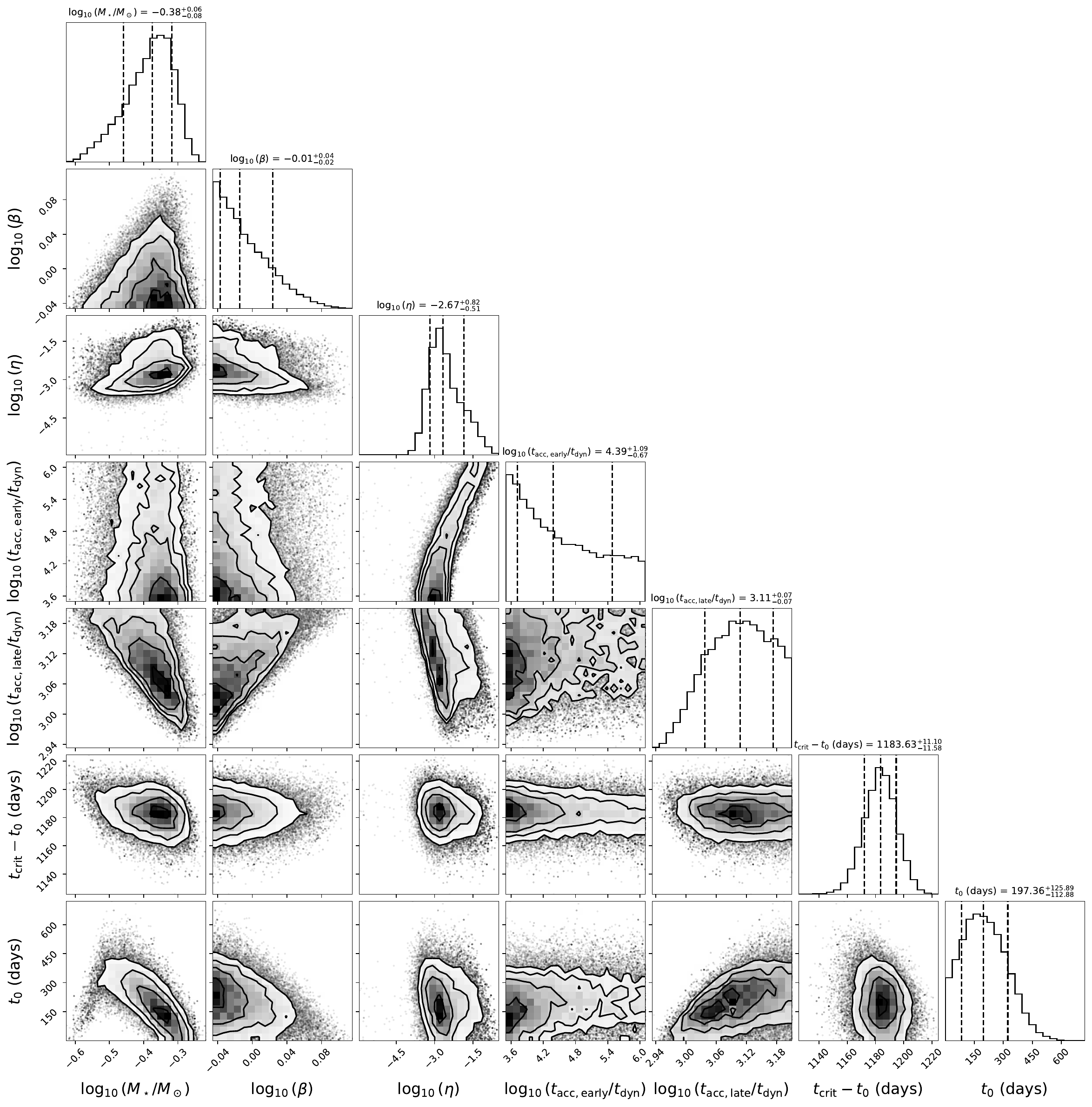} 
\caption{EP240222a model fit corner plot. Quoted values and black dashed lines denote the posterior medians and 1$\sigma$ (68\%) credible intervals. Darker 2D regions indicate higher probability density.}
\label{fig:corner}
\end{figure*}

We retained 77 walkers after pruning low acceptance rates ($<0.015$). We discarded the initial $60,000$ steps (burn-in) for convergence and thinned chains by a factor of $50$ to reduce autocorrelation.

Visual inspection identified two modes. We discarded solutions with $\log_{10}(t_{\mr{acc,late}}/t_{\mr{dyn,c}}) > 3.21$ as nonphysical. Their extended flat light curves before the turning point contradict the growing disk mass (rising luminosity) needed to trigger stream-disk interaction. Following extraction of self-consistent $t_{\mr{acc,early}}$ (detailed in \S\ref{subsec_stream_disk}), Fig.~\ref{fig:corner} displays the final corner plot.

The physically plausible posteriors reproduce the peculiar light curve with natural parameters, avoiding extreme values. $M_*$, $\beta$, $t_{\mr{acc,late}}$, and $t_{\mr{crit}}-t_0$ are well-constrained; $\eta$ and $t_0$ less so; $t_{\mr{acc,early}}$ yields only an upper limit.

This can be understood physically. Multi-stage data constrain $M_*$ and $\beta$, while the clear turning point locks $t_{\mr{crit}}-t_0$. This determines the plateau-onset disk mass ($\sim M_{\mr{s}}(t_{\mr{crit}})$), which, combined with precise plateau and decline data, constrains $t_{\mr{acc,late}}$. Conversely, sparse, high-uncertainty data of the early stage yield weaker constraints on $\eta$, $t_0$, and $t_{\mr{acc,early}}$.

Figure~\ref{fig:mcmc} compares data [eROSITA (red), LEIA (orange), EP-FXT (olive), NICER (blue), XMM-Newton (black)] with median-parameter model (red solid), point-wise median model (black dashed), and the corresponding $1\sigma$ (68\%) credible interval (gray shade). The poorer slow-rise fit results from sparse, high-uncertainty early data, physical perturbations (e.g., from the main stream and Lense-Thirring precession), and MCMC bias favoring denser, precise late data.

In conclusion, the good constraints on the parameters and the fact that the fitting result fits the data well validate our model, indicating EP240222a involved a $M_* \approx 0.4~M_\odot$ MS star with $\beta \approx 1.0$.

\section{Discussion} \label{sec_dis}

\subsection{The Mass of the Main Stream} \label{subsec_main_stream_mass}

We justify the assumptions for $M_s(t)$ (Eq.~\ref{eq:ms_t}): full tail incorporation and negligible main stream mass loss.

First, the tail's low momentum flux relative to the main stream~\citep{Bonnerot2017} ensures full assimilation, creating a smooth merger without significant energy or angular momentum outliers that could generate separate inflows or outflows. Second, the main stream remains coherent after nozzle shocks~\citep{Bonnerot2022, Hu2025, Noah2025}, with negligible shedding.

Finally, regarding self-intersection losses (inflow and outflow; \citealt{Lu2020,Huang2023}), simulations show small outflows for $M_{\mr{BH}} \lesssim \text{a few} \times 10^6~M_\odot$~\citep{Lu2020}. For IMBH-TDEs, smaller apsidal precession angle yields shallower, lower-velocity apocenter collisions; this "gentle" interaction minimizes outliers and mass loss. This is validated self-consistently: our fit requires minimal accumulated inflow mass to the precursor disk ($\lesssim 0.02 M_\mr{fb}$, using the median parameters). Since inflow and outflow masses should be comparable, this implies a similarly small outflow, matching simulations and bolstering our framework's reliability.

\subsection{Consistency with Optical Light Curves} \label{subsec_opt_light_curve}

While constructed for the unique, long-term X-ray light curve, our model must self-consistently explain the optical emission as a crucial test.

\begin{figure*} 
\centering
\includegraphics[width=\textwidth]{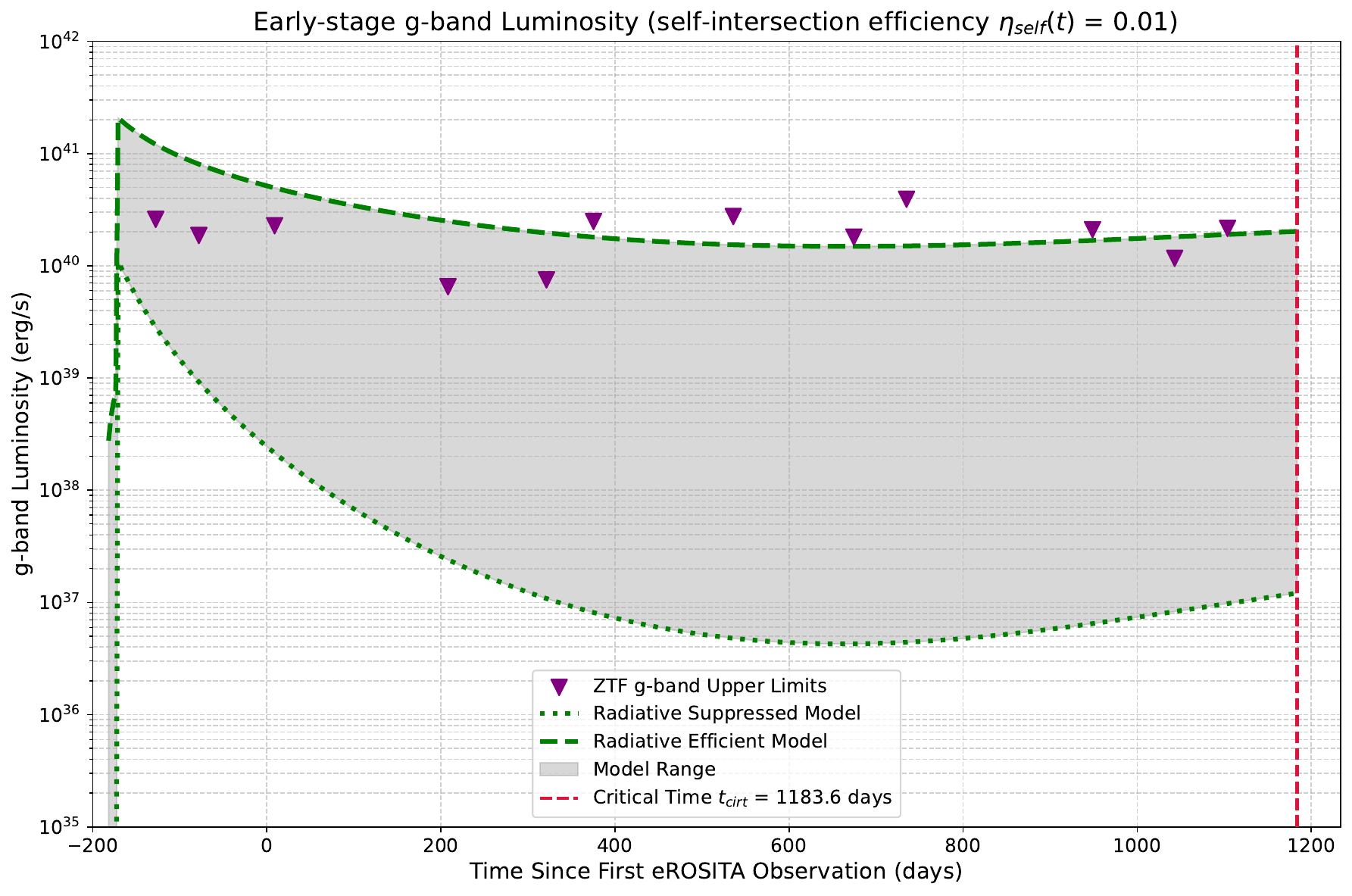} 
\caption{Early optical models vs. ZTF upper limits (purple arrows, 90\% CL). Green lines show lower (persistent suppression) and upper (fully efficient radiation) bounds for $\eta_{\mr{self}}(t)=0.01$, utilizing median parameters in Fig.~\ref{fig:corner}. ZTF data are consistent with the expectation that the true light curve evolves from the lower to the upper bounds, constraining $\eta_{\mr{self}}(t) \lesssim 0.01$. This implies highly inefficient self-intersection and/or inaccuracies in our simple thermal correction. In either case, we conclude inherently faint optical emission. The upper limits, all preceding fitted $t_{\mr{crit}}$ (red dashed), support our model as well.}
\label{fig:early_optical}
\end{figure*}

\subsubsection{Early Upper Limit} \label{sub_sub_sec_early_opt}
Early non-detections ($L_{\mr{g,early}} \lesssim 10^{40}$~\lum), all prior to $t_\mr{crit}$, align with our model. Furthermore, their faintness is expected from the four primary emission channels.

(1) {\bf X-ray reprocessing by the stream} is intrinsically faint. Even assuming a reprocessing efficiency similar to later times yields only $L_g \sim 10^{39}$~\lum, and the true efficiency is likely even lower because most X-ray photons scatter off the stream's surface rather than being reprocessed.

(2) {\bf Nozzle shock emission} is inherently limited, with peak luminosity ~\citep{Bonnerot2021,Bonnerot2022}:
\begin{equation}
\label{eq:L_peak_noz}
\begin{split}
L_{\mr{peak,noz}} & \simeq \frac{\eta_{\mr{noz}} \left(\frac{1}{2} M_{\mr{fb}} v_{\mr{z,max}}^2\right)}{t_{\mr{fb}}} \\
                 & \simeq 8.5 \times 10^{39}~\eta_{\mr{noz}} \beta^{2} M_5^{-1/2} \frac{M_\mr{fb}}{M_*/2} m_{*}^{3} r_{*}^{-5/2}~\mr{erg~s^{-1}} \\
                 & \quad \times \begin{cases}
                     \beta^{3}, &\beta \lesssim \beta_{\mr{d}} \\
                     1, & \beta \gtrsim \beta_{\mr{d}},
                 \end{cases}
\end{split}
\end{equation}
where the efficiency $\eta_{\mr{noz}} \ll 1$ due to high optical depths. Combined with shock weakening after a few orbits (\S\ref{subsec_ssc}), this renders early emission (including potential X-rays) negligible.

(3) {\bf SSC and tail-main stream interaction} dominate the optical luminosity during the slow-rising stage, modeled as\footnote{Tail-main stream interaction starts at $2.5~t_\mr{fb}$ (\S\ref{subsec_ini}). Minor differences between idealized and actual fallback parameters are neglected.}
\begin{equation}
\label{eq:L_slow_rise}
L_{\mr{g,early}}(t) \simeq \frac{M_{\mr{s}}(t) \dot{E}(t) + \dot{M}_{\mr{fb}}(t)\left[E_0-E_{\mr{fb}}(t)\right]}{K_\mr{g}(t)},
\end{equation}
where $E_{\mr{fb}}(t) = \left[\pi^2G^2 M_{\mr{BH}}^2 /({2t^2})\right]^{1/3}$ is the specific binding energy of the tail falling back at $t$. We model the emitting region as a uniform sphere and utilize the blackbody approximation due to its high optical depth ($\tau_{\mr{es}} \simeq \kappa_{\mr{es}} M_\mr{fb}/(4 \pi a_0^2) \approx 2.7\times10^{3}$). The temperature $T_{\mr{opt,early}}(t)$ is set by $L_{\mr{opt,early}}(t) \simeq 4\pi a_0^2 \sigma T_{\mr{opt,early}}^4(t)$, giving $K_\mr{g}(t) \simeq \sigma T_{\mr{opt,early}}^4(t) / \left[\pi \nu_\mr{g,early} B_\mr{g,early}(T_\mr{opt,early}(t))\right]$, where $B_\nu(T)$ is the Planck function.

Radiation trapping initially suppresses luminosity to $\approx 0.0065$ (Eq.~17 in \citealt{Bonnerot2021}), becoming efficient via stream expansion after a few $t_\mr{fb}$ (\S\ref{subsec_ssc}). Figure~\ref{fig:early_optical} shows lower (persistent suppression) and upper (fully efficient radiation) bounds for $\eta_{\mr{self}}(t)=0.01$\footnote{We assume $\eta_{\mr{self}}(t)$ is constant for order-of-magnitude estimation, as its total change (though decreasing; \S\ref{subsec_ssc}) is less than an order of magnitude.} utilizing median parameters (Fig.~\ref{fig:corner}) alongside ZTF upper limits. The limits match the expectation that the true light curve evolves from lower to upper bounds, constraining $\eta_{\mr{self}}(t) \lesssim 0.01$. This implies highly inefficient self-intersection and/or inaccuracies in our simple thermal correction. In either case, we conclude inherently faint optical emission. A detailed physical explanation requires future theoretical and numerical studies.

In conclusion, all early optical channels are inherently faint for EP240222a, naturally explaining $L_{\mr{g,early}} \lesssim 10^{40}$~\lum. Furthermore, we predict that IMBH-TDEs of MS stars should, in their early stages, exhibit a faint, years-long optical light curve characterized by a slow rise/quasi-plateau. This prediction offers a key observational signature for identifying such events in future time-domain surveys.

\subsubsection{Plateau and Decline Stage} \label{Plateau_and_Decline}
During these stages, optical light curves show two distinct behaviors compared to the X-rays, supporting the reprocessing scenario. First, a slower decline because of the smoothing effect of reprocessing and photospheric cooling shifting emission redward (see Eq.~9 in \citealp{Chen2018}). Second, the plateau-to-decline turnover is delayed due to the reprocessing time.

Despite robust reprocessing evidence, optical luminosity is a factor of $\sim 200$ fainter than X-rays, contrary to naive expectations. We attribute this to three effects.

First, viewing angle~\citep{Dai2015,Qiao2025}: face-on orientation exposes the central X-ray source. Second, reprocessing SED peaking in FUV/EUV: due to hotter reprocessing photospheres for IMBH-TDEs~\citep{Chen2018}, the optical band captures only the faint spectral tail (Fig.~S12 in \citealp{Jin2025EP240222a}). Third, radiative transfer complexity: sophisticated {\tt{Sedona}} simulations~\citep{Jin2025EP240222a} yield $L_{\rm opt} \sim 10^{41}$~\lum, matching observations and underscoring limitations of simpler models.

\subsection{Rise Timescale for IMBH-TDEs} \label{subsec_rise_timescale}

While our model successfully explain the unique prolonged rise of EP240222a, not all IMBH-TDEs necessarily exhibit this characteristic. Rapid circularization can occur for an IMBH in the disruption of a white dwarf (WD), a type of TDE unique to IMBHs~\citep{Gezari2021}. Extremely small WD radii enhance self-intersection efficiency by several orders of magnitude ($\epsilon_{\mr{self}} \propto R_*^{-2}$), potentially driving single-orbit circularization. Here, negligible circularization timescale leads to prompt massive disk formation and Eddington accretion, reverting to the \citet{Chen2018} regime. 

Further considering the crucial role of the viewing angle~\citep{Dai2018, Qiao2025}, the IMBH-WD TDEs can produce two distinct types of transients. The first is the "off-axis" case, resembling typical optical/UV TDEs but with a faster rise ($\sim t_\mr{fb} \sim$ minutes-days; Eq.~\ref{eq:t_fb}) and a longer plateau. Depending on the specific viewing angle, X-rays around the optical/UV peak can be either bright or faint, and their onset is often delayed. In the second case, "on-axis" observers see rapidly rising ($\sim t_\mr{fb} \sim$ minutes-days; Eq.~\ref{eq:t_fb}), thermal X-rays through the optically thin funnel, featuring similar optical/UV evolution, and potentially accompanied by a brighter and harder jet component enhanced by relativistic beaming. This scenario may explain exotic transients such as EP250702a~\citep{Li2025}. We caution, however, that IMBH-WD TDEs involve complex physics~\citep{Maguire2020} requiring comprehensive future theoretical, numerical, and observational study.

Finally, our model explains the rarity of EP240222a-like two-stage rise: most known TDEs host massive BHs ($M_{\mr{BH}} \gtrsim 10^6M_\odot$). Their higher $\epsilon_{\mr{self}}$ shortens the "slow-rising stage", causing initial circularization driven by the SSC and final, rapid circularization from the stream-disk interaction to occur almost concurrently. Furthermore, strong and rapid energy release blurs the distinction between the precursor disk and the main stream, merging two stages into a single, rapid circularization event. This would produce a light curve with a single, fast rise, masking the separate physical stages that were clearly delineated in EP240222a.

\subsection{The Critical Moment for Stream-disk Interaction} \label{subsec_stream_disk}

In our model, we treat the critical time ($t_{\mr{crit}}-t_0$) for the onset of strong stream-disk interaction as a free parameter. Here, we perform a self-consistency check to see if our fitting results are physically plausible. 

Strong interaction triggers when stream and disk momentum fluxes match. Since the main stream is on a nearly parabolic trajectory, its velocity is approximately $\sqrt{2}$ times the local Keplerian velocity of the circular disk. The condition therefore simplifies to mass fluxes, scaled by this velocity factor and the disk aspect ratio:
\begin{equation}
\label{eq:stream_disk}
\sqrt{2}\dot{M}_{\mr{s}}(t_\mr{crit}) \left(\frac{H}{R}\right)_{\mr{d,early}}\simeq \dot{M}_{\mr{d}}(t_\mr{crit}).
\end{equation}
The term $(H/R)_\mr{d,early}$ accounts for cross-sectional differences: the main stream is thick ($(H/R)_\mr{s} \sim 1$; \S\ref{subsec_ssc}), while the precursor disk is likely thinner ($(H/R)_{\mr{d,early}} \ll 1$).

Mass fluxes are $\dot{M}_{\mr{s}}(t_\mr{crit})\simeq M_{\mr{s}}(t_\mr{crit}) / t_{\mr{s,N}}(t_\mr{crit})$ and $\dot{M}_{\mr{d}}(t_\mr{crit}) \simeq M_{\mr{d}}(t_\mr{crit}) / t_{\mr{orb,c}}$, where the orbital timescale for a circular orbit at $R_\mr{c}$ is $t_{\mr{orb,c}} = 2\pi\sqrt{R_{\mr c}^3/(G M_{\mr {BH}})}$.

By combining these relations, we can derive:
\begin{equation}
\label{eq:self_consistency_relation}
\left(\frac{H}{R}\right)_{\mr{d,early}}^{3} \alpha_{\mr{early}} \simeq \frac{t_\mr{s,N}(t_\mr{crit}) L_\mr{bol}(t_\mr{crit})}{2\sqrt{2}\pi \eta_{\mr{mis}} M_{\mr{s}}(t_\mr{crit}) c^2},
\end{equation}
assuming $(H/R)_{\mr{s}} \approx 1$ at the circularization radius and substituting Eqs.~\ref{eq:t_acc} and \ref{eq:l_bol}.

This equation constrains precursor disk properties using our fitting results: $(H/R)_\mr{d,early}^{3} \alpha_{\mr{early}}$ ranges from $\approx 1.5 \times 10^{-6}$ (maximally prograde; $\eta_{\mr{mis}} \approx 0.234$) to $\approx 9.1 \times 10^{-6}$ (maximally retrograde; $\eta_{\mr{mis}} \approx 0.038$). By applying limits $0.001 < (H/R)_{\mr{d,early}} < 2$ and $\alpha_{\mr{early}} < 0.4$ (\S\ref{subsec_ssc}), we can further constrain the $\log_{10}(t_{\mr{acc, early}}/t_{\mr{dyn,c}})$ posterior to $[3.5, 6.1]$, leaving other parameters largely unchanged. This implies a plausible relatively thin, inefficiently accreting precursor disk (e.g., $\alpha_{\mr{early}} \sim 0.02$, $(H/R)_{\mr{d,early}} \sim 0.04$). This successful self-consistency validates our model, though detailed study of the stream-disk interaction requires future simulations.

\section{Conclusion} \label{sec_con}
We have developed a five-stage, semi-phenomenological model that successfully reproduces the complex multi-wavelength light curves of the IMBH-TDE EP240222a, the first such event captured in real-time with multi-wavelength observations and spectroscopic confirmation, and one whose light curves largely deviate from previous expectations for IMBH-TDEs. Our model reveals key physical processes such as inefficient circularization, delayed stream-disk interaction, and reprocessing. It successfully constrains EP240222a to be the disruption of a $M_* \approx 0.4~M_\odot$ MS star with a penetration factor $\beta \approx 1.0$. Furthermore, our model illustrates distinctive features of IMBH-TDEs that can inform future searches for such events.

Specifically, our proposed model delineates the evolution of the IMBH-TDE, as exemplified by EP240222a, into the following stages. During the initial and slow-rising stages, we identify an inefficient circularization process. The inflow caused by a succession of self-crossings of the main stream forms a precursor disk, accounting for the low-luminosity X-ray emission with a long rise time. Meanwhile, the faintness of the early optical emission (as constrained by upper limits) is primarily due to a combination of inefficient energy dissipation, photon diffusion, and a potentially large bolometric correction. The subsequent fast-rising stage is triggered by a delayed stream-disk interaction, once momentum flux matching occurs between the disk and the main stream. This interaction completes the circularization of the main stream, rapidly increases the disk mass, and causes the sharp rise in accretion rate and luminosity. The following X-ray plateau is shown to be driven by radiative suppression from super-Eddington accretion. In fact, our model reveals it as a "quasi-plateau" with a slow, linear decline. The simultaneous optical plateau arises from a super-Eddington outflow that reprocesses the central X-ray emission, while the unobscured nature of the X-rays is explained by our line of sight looking down an optically thin funnel. Finally, the decline stage marks the transition to sub-Eddington accretion. The X-ray emission evolves from a linear decay toward a power-law decay, while the delayed and slower optical decline is consistent with a reprocessing origin, confirming reprocessing as the dominant mechanism for late-time optical emission.

Our findings offer crucial insights for future searches for IMBH-TDEs. Specifically, EP240222a can serve as a prototype for the tidal disruption of MS stars by IMBHs. A key observational signature for such events is a faint, slow-rising early X-ray phase (on timescales of years), followed by an Eddington-luminosity plateau. In the UV/optical bands, this sequence manifests as a faint, year-long rise/quasi-plateau corresponding to the early X-ray stage, followed by a brighter plateau that tracks the super-Eddington stage. In contrast, we predict a distinct signature for the tidal disruption of a WD by an IMBH: a fast-rising flare (on timescales of minutes to days) in X-rays (thermal component) and/or UV/optical, followed by an Eddington-luminosity plateau lasting for months. Although IMBH-TDEs could be inherently rare and observationally diverse, their systematic discovery is now a promising prospect in the new era of time-domain astronomy. This is driven by high-cadence X-ray surveys like EP and deep optical surveys such as the Wide Field Survey Telescope (WFST; \citealt{WFST}) and the Legacy Survey of Space and Time (LSST; \citealt{LSST}).

\acknowledgments
W.L. thanks the organizers and lecturers of the 2024 WFST Summer School, especially Dr. Tinggui Wang for the lectures on TDEs and Dr. Hua Feng for the lectures on accretion; the organizers and lecturers of the TDE Full-Process Simulation Workshop for the valuable discussions on TDEs; Xiaoyu Zhuang for suggestions on the schematic diagram; and Yue Zhang for advice on the MCMC. Valuable discussions with USTC colleagues and helpful questions from the Steward Science Symposium are gratefully acknowledged. This work is supported by the Strategic Priority Research Program of the Chinese Academy of Sciences (XDB0550200), the National Natural Science Foundation of China (grants 12522303,12192221,12393814), and the National College Students' Innovative Entrepreneurial Training Plan Program. This material is based upon High Performance Computing (HPC) resources supported by the University of Arizona TRIF, UITS, and Research, Innovation, and Impact (RII) and maintained by the UArizona Research Technologies department.

\bibliography{ref}{}
\bibliographystyle{aasjournal}

\end{document}